\documentclass[showpacs,preprint,preprintnumbers,amsmath,amssymb,nofootinbib]{revtex4-1}

\usepackage{color}
\definecolor{Blue}{rgb}{0.3,0.3,0.9}
\usepackage{graphicx}
\usepackage{subfigure}
\usepackage{dcolumn}
\usepackage{bm}
\usepackage{epsfig}
\usepackage{amsfonts}

\begin{document}

\title{Renormalized entropy for  one dimensional discrete maps: 
periodic and quasi-periodic route to chaos and their robustness}

\author{O. Afsar$^{1,}$}
\email{ozgur.afsar@ege.edu.tr}
\author{G. B. Bagci$^{1,}$}
\email{baris.bagci@ege.edu.tr}
\author{U. Tirnakli$^{1,2,}$}
\email{ugur.tirnakli@ege.edu.tr} \affiliation{
$^1$Department of Physics, Faculty of Science, Ege University, 35100 
Izmir, Turkey\\
$^2$Division of Statistical Mechanics and Complexity, Institute of Theoretical 
and Applied Physics (ITAP) Kaygiseki Mevkii, 48740 Turunc, Mugla, Turkey
}

\date{\today}

\begin{abstract}

We apply renormalized entropy as a complexity measure to the
logistic and sine-circle maps. In the case of logistic map,
renormalized entropy decreases (increases) until the accumulation
point (after the accumulation point up to the most chaotic state)
as a sign of increasing (decreasing) degree of order in all the
investigated periodic windows, namely, period-$2$, $3$, and $5$,
thereby proving the robustness of this complexity measure. This
observed change in the renormalized entropy is adequate, since the
bifurcations are exhibited before the accumulation point, after
which the band-merging, in opposition to the bifurcations, is
exhibited. In addition to the precise detection of the
accumulation points in all these windows, it is shown that the
renormalized entropy can detect the self-similar windows in the
chaotic regime by exhibiting abrupt changes in its values.
Regarding the sine-circle map, we observe that the renormalized
entropy detects also the quasi-periodic regimes by showing
oscillatory behavior particularly in these regimes. Moreover, the
oscillatory regime of the renormalized entropy corresponds to a
larger interval of the nonlinearity parameter of the sine-circle
map as the value of the frequency ratio parameter reaches the
critical value, at which the winding ratio attains the golden
mean.

\end{abstract}


\maketitle

\section{Introduction}

With the advent of the science of complexity, numerous complexity
measures have been proposed. These measures can be grouped mainly
in two categories: first group follows the route of constructing
the shortest computer program corresponding to a given string. The
well-known Kolmogorov-Chaitin \cite{Kolmogorov, Chaitin} and
logical depth \cite{Bennett} measures fall into this category. The
second group follows the information-theoretic approaches whose
main examples can be given as effective complexity, thermodynamic
depth \cite{Lloyd}, Shiner-Davison-Landsberg \cite{Shiner} and
L\`{o}pez-Ruiz-Mancini-Calbet measures \cite{Calbet}.

The information-theoretic approaches are founded on the main idea
of multiplying a measure of order by that of disorder. In this
sense, these approaches rely heavily on the definitions of entropy
as a measure of order/disorder. For example,
Shiner-Davison-Landsberg \cite{Shiner} uses Boltmann-Gibbs-Shannon
(BGS) entropy of the physical system in its definition, while
thermodynamic depth as a complexity measure considers the entropy
of the ensemble focusing on the entire history of the system under
investigation \cite{Lloyd}.

However, to the best of our knowledge, none of these complexity
measures have been constructed particularly to deal with the
non-equilibrium stationary states resulting from the external
influence of a field. Such a complexity measure has been recently
introduced by Saparin et al. and applied to logistic map
\cite{Kurths1}, heart rate variability \cite{Kurths2,Kurths3} and
to the analysis of electroencephalograms of epilepsy patients
\cite{Kurths4}. This new complexity measure is an
information-theoretic one and is called renormalized entropy,
historically originating from Klimontovich's S-theorem (The letter
S here stands for the self-organization)
\cite{Klimontovich1,Klimontovich2, Klimontovich3, Klimontovich4,
Klimontovich5}. The renormalized entropy is theoretically
equivalent to negative relative entropy between a reference
distribution and any other distribution obtained either
analytically or numerically through a time-series analysis
\cite{Kopitzki,Grassberger}. However, it is not only associated
with the relative entropy, since an additional procedure of
renormalization is also introduced. The process of renormalization
is used in order to compare two states by equating their mean
energies so that non-equilibrium stationary states attain the same
mean energy, thereby mimicking the ordinary closed system
formalism. Combining renormalization and the relative entropy, the
renormalized entropy decreases as the control parameter increases,
indicating the relative degree of order in the system as first
suggested by Haken \cite{Haken} in the context of
self-organization.

On the other hand, another related issue in the context of
self-organization is the existence of bifurcations often observed
in nonlinear dynamical systems: the systems possessing a stable
fixed point become unstable as they recede away from this stable
fixed point as a result of increasing nonlinear effects. These
systems eventually pave their way to the new stable branches
through bifurcations. For the system away from the stable fixed
point, this process continues until the system sets in a new
stationary state, thereby increasing its order as a signature of
self-organization due to the non-linearity, dissipation and the
non-equilibrium exhibited by the system (see the detailed
discussion by Nicolis and Prigogine in Ref. \cite{Prigogini}).
Physical processes such as Rayleigh-Benard 
\citep{Rayleigh1916,Rayleigh-Benard1985}, 
Taylor instability \citep{Taylor1923} experiments, 
bacterial \citep{Benjacob1994} and {\it{Dictyostelium discoideum}} 
\citep{Palsson1996} colonies fall into the aforementioned category.
These systems, being open due to exchange of energy and/or 
matter with its surroundings, can not be analyzed in terms of usual H-theorem,
since it is valid only for isolated systems.
Therefore, Prigogine proposed a more general form of the
second law i.e. $\Delta{S}=\Delta_i{S}+\Delta_e{S}$
where $\Delta_i{S}$ is the entropy produced inside the system
and $\Delta_e{S}$ is the transfer of entropy across the boundaries of the system. 
In this more general setting, $\Delta_i{S}\ge 0$ whereas $\Delta_e{S}$ can be
positive or negative depending on the flow of energy across the boundaries
of the system. The overall sign of $\Delta{S}$ is determined by the
interplay between $\Delta_i{S}$ and $\Delta_e{S}$. 
All the aforementioned models of self-organization requires
$\Delta{S} < 0$ until the system sets in the stationary state
corresponding to the most ordered pattern. 

The paper is organized as follows. In Section 2, calculation methods of the renormalized
entropy is given. In section 3, this complexity measure is applied to discrete maps possessing different universality classes i.e. to the logistic map which has periodic route to chaos
and its self similar windows like period 3 and period 5 to show its robustness
and to sine-circle map which has periodic and quasi-periodic route to chaos. 
Finally, we discuss our results and
compare the two different routes to chaos in terms of the renormalized entropy.

\section{Method}
\subsection{An overview for the renormalized entropy}

Let us consider a generic dynamical map $X_{t+1}=G(X_{t},a)$ where
$t$ and $a$ denote the number of iterations and the corresponding
control parameter of the dynamical map, respectively. A small
change in the value of the control parameter, $\Delta{a}$, yields
two normalized distributions i.e. $f(X,a)=f_{0}(X)$ and
$f(X,(a+\Delta{a}))=f_{1}(X)$. 
The corresponding Shannon entropies read
\begin{equation}
S\left ({f_{0}(X)}\right )=-\int f_{0}(X)\, \ln f_{0}(X)dX \,\,\,\, , 
\,\,\,\,  S\left ({f_{1}(X)}\right )=-\int f_{1}(X)\, \ln f_{1}(X)dX \, .
\label{Shannon}
\end{equation}
Let us now assume that the system i.e. dynamical map under
scrutiny evolves in such a manner that the state with index $1$ is
evolved to through increasing order, namely, the system becomes
self-organized as the control parameter increases. Then, setting
equilibrium temperature
$T_{eq}=1$, the normalized Boltzmann-Gibbs distribution
reads

\begin{equation}
f_{0}(X)=\exp \left [-H_{eff}(X)\right ]
\label{Boltzmann}
\end{equation}
with the following effective energy

\begin{equation}
H_{eff}=-\ln f_{0}(X).
\end{equation}
S-theorem by Klimontovich equates the effective energies of the
concomitant states i.e. renormalizes the states in order to apply
H-theorem of Boltzmann to open systems, implying $\Delta_e{S}=0$ in the second 
law formulation of Prigogine \citep{Nobel} and $f_{0}(X) \rightarrow \tilde{f}_{0}(X)$
to compensate for the mean energy difference i.e. turning open system
into a closed one. 
Denoting the renormalized
state by $tilde$, one can write

\begin{equation}
\tilde{f}_{0}(X)=C\,\left [ f_{0}(X) \right ]^{\beta_{eff}}=
C\,\exp \left [ \frac{-H_{eff}(X)}{T_{eff}} \right]
\label{renormalization}
\end{equation}
where $C$ is the normalization constant. To check whether any heat
intake occurs during self-organization, to form spatially more ordered
patterns, through dissipation as a
result of interaction with the environment, $T_{eff}$ is
calculated from the equality of mean energies

\begin{equation}
\int \tilde{f}_{0}(X)\, \ln f_{0}(X)dX=\int f_{1}(X)\, \ln f_{0}(X)dX \,.
\label{energyequality}
\end{equation}

If heat intake is needed for the process of self-organization such
as Rayleigh-Benard convection resulting in spatially ordered
hexagonal patterns at the stationary state
\citep{Rayleigh1916,Rayleigh-Benard1985}, 
then one expects, comparing the
equilibrium and non-equilibrium states, $T_{eff} > 1$, since
$T_{eq}=1$. Otherwise, we deduce that our assumption
regarding the more disorderliness of the initial distribution is
not correct, indicating the second distribution to be more
disordered. Therefore, when this is the case, one renormalizes the
second distribution. The measure of relative degree of order for
these compared states can then be given as the difference of
entropies

\begin{equation}
\Delta \tilde{S}=S\left ({f_{1}(X)}\right )-S \left({\tilde{f}_{0}(X)}\right )
\label{relatifentropy}
\end{equation}
which is called renormalized entropy \citep{Anishchenko1994}.

\subsection{Calculation of the spectrum}

We use the spectral intensities
$P_0(w)$ and $P_1(w)$ averaged over $M$ periodograms based on 
the multiplication of the 
Fourier and the inverse Fourier transformation
of the time series of $X_t$ in the frequency domain $w$, 
instead of the density distributions $f_{0}(x)$ and $f_{1}(x)$ 
in Eqs.~[\ref{Shannon}-\ref{relatifentropy}], respectively
so that
\begin{equation}
P(w)=\frac{1}{\ell\,\Delta{t}}
\left |\Delta{t}\,\sum_{n=0}^{\ell-1} x_n\,\exp(-i2{\pi}wn\Delta{t})\right |^2 \;,
\label{spectral}
\end{equation}
where 
the sequence $X_n$ is a sufficiently long series of values
that can be obtained by iterating a mapping by sampling equidistant points
and
$\ell$ is the length of the samples in every  periodogram \cite{wessel1},
satisfying $P(w)>0$ and $\sum{P(w)=1}$.
Such a Fourier spectra eliminate the 'zeros' in the distributions
and detect  different regimes of a deterministic dynamical system.


\section{Application of the renormalized entropy}

To start with, let us consider a $d$-dimensional mapping of the
form
\begin{equation}
X_{t+1}=G (X_t)  \;
\label{mapping}
\end{equation}
on some $d$-dimensional phase space $X$. We can numerically
generate data from such a mapping equation
that describes dynamics of a specific dynamical system.
We particularly focus on two examples. 
One is the logistic map given as

\begin{equation}
X_{t+1}=4\,a\, X_t\,(1-X_t)
\label{logisticmap}
\end{equation}
where $a\in[0,4]$ is a control parameter and the map is confined
to the interval $X\in[0,1]$, which exhibits periodic route to chaos. 
For $a=4$, the system is
(semi)conjugated to a Bernoulli shift and strongly mixing.
The other is the sine-circle map given as

\begin{equation}
\theta_{t+1}=\theta_t +\Omega -\frac{K}{2\pi}\sin{(2\pi\theta_t)}
~~~ \mod(1) \;\;
\label{circlemap}
\end{equation}
where $0\leq\theta_t<1$ is a point on a circle and parameter $K$
(with $K>0$) is a measure of the strength of the nonlinearity,
whichs exhibit periodic route to chaos. 
It describes dynamical systems possessing a natural frequency
$\omega_1$ which are driven by an external force of frequency
$\omega_2$ ($\Omega =\omega_1/\omega_2$ is the bare winding number
or frequency-ratio parameter) and belongs to the same universality
class of the forced Rayleigh-Benard convection \cite{jensen}.
Winding number for this map is defined to be the limit of the
ratio

\begin{equation}
W =\lim_{t\rightarrow\infty} \frac{(\theta_{t}-\theta_0)}{t} \;\; ,
\label{winding}
\end{equation}
where $(\theta_{t}-\theta_0)$ is the angular distance travelled
after $t$ iterations of the map function.
To increase the degree of irrationality of the system, 
one could use frequency ratio parameter $\Omega$
corresponding to winding number $W$
that approaches the golden mean gradually by following the sequence
of ratios of the Fibonacci numbers (${F_n}/{F_{n+1}}$).
The map is monotonic and
invertible (non-monotonic and non-invertible) for $K<1$ ($K>1$)
and develops a cubic inflexion point at $\theta=0$ for $K=1$.

Considering technical details, it is important to emphasize that
we added Gaussian white noise into systems in
Eqs.~(\ref{logisticmap}) and (\ref{circlemap}) 
with a small intensity $(D_B=10^{-14})$, which is called
basic noise, to every state. The low intensity of the noise is chosen 
so as not to influence the dynamics of
these systems. This procedure enables 
a continuous spectral distribution $P(w)$.
After 65536 transients, we generated the discrete sequences with
the length of $4096\times 18$ points separately obtained from Eq. (9) and
(10). Finally, the spectrum of 18 shifted windows 
of 4096 samples $(\Delta{t}=0.5~s)$ is estimated and averaged.

\subsection{Renormalized entropy and its robustness at
different windows for the logistic map: periodic route to chaos}

One of the dynamical systems possessing bifurcation properties
can be cited as the logistic map, which
moves from a unique stable fixed point to the critical
accumulation point possessing infinite periods through the
periodic doubling route. Having reached the
critical accumulation point, the system enters into the chaotic
regime and from this point on it moves under the influence of
chaotic band merging (i.e., inverse period-doubling). It should
also be remarked that periodic windows with different periods but
albeit self-similar structures are also found in this chaotic
regime.

We now present the results concerning the behavior of the
renormalized entropy and the bifurcation properties in the regions of
period $2$, $3$ and $5$. Before proceeding further, we note that the renormalized 
entropy has already been applied to the logistic map for period-2 window 
by Saparin \textit{et al.} in Ref. \cite{Kurths1}. 
Our aim in this section is to investigate all other self-similar windows 
to check whether the renormalized entropy behaves consistently as a 
complexity measure, i.e., to check its robustness. Fig.~1a in particular 
shows the behavior of the renormalized entropy in the period 2 region where 
control parameter $a$ lies between $3$ and $4$. In this region, relative
degree of order increases until the period accumulation point
i.e., $a_{c}^{\left( 2\right) }\sim 3.569$ as the system evolves
from the equilibrium state to a new stationary state in accordance
with the self-organization process. As a signature of the
detection of the self-organization in this region then, the
relative entropy monotonically decreases as expected. After the
period $2$ accumulation point onward until the most chaotic state
with $a=4$, the relative degree of order decreases, since
band-merging (as opposed to bifurcation in the previous region) is
exhibited by the system in this region. Therefore, the
renormalized entropy increases in the aforementioned region in a
non-monotonic manner. To sum up, for an open non-equilibrium
dynamical system which approaches the stationary state through
period-doubling and recedes away from the stationary state by
means of band-merging, the behavior of the renormalized entropy
conforms to the dictum of Prigogine i.e., order out of chaos. In
Fig.~1b, we zoom in the period $3$ window i.e., the largest
self-similar window in the period $2$ region. The control
parameter values for this window are confined in the interval
between $3.832$ and $3.856$. Similarly to the period $2$ window,
the relative degree of order monotonically increases up to the
period accumulation point $a_{c}^{\left( 3\right) }\sim 3.849$,
and thereafter decreases non-monotonically. Accordingly, the
relative entropy decreases up to the accumulation point, 
and begins to increase after the period accumulation
point. It is worth noting the sudden, unexpected changes in the
values of the renormalized entropy which signal the existence of
the self-similar windows in the chaotic region. Fig.~1c shows the
behavior of the renormalized entropy in the period $5$ window,
which is one of the self-similar windows in the logistic map. Due
to the self-similarity, a behavior similar to the one in period
$3$ is exhibited by the renormalized entropy: it decreases almost
up to $a_{c}^{\left( 5\right) }\sim 3.743$, and then begins to
increase in accordance with the decrease in the relative degree of
order. Finally, it is interesting to observe the turns in the
relative degree of order for each of the three period accumulation
points representing the stationary state of a non-equilibrium
dynamical system possessing inherent fractal structure.

\begin{figure}
\begin{center}
\includegraphics*[height=6.5cm]{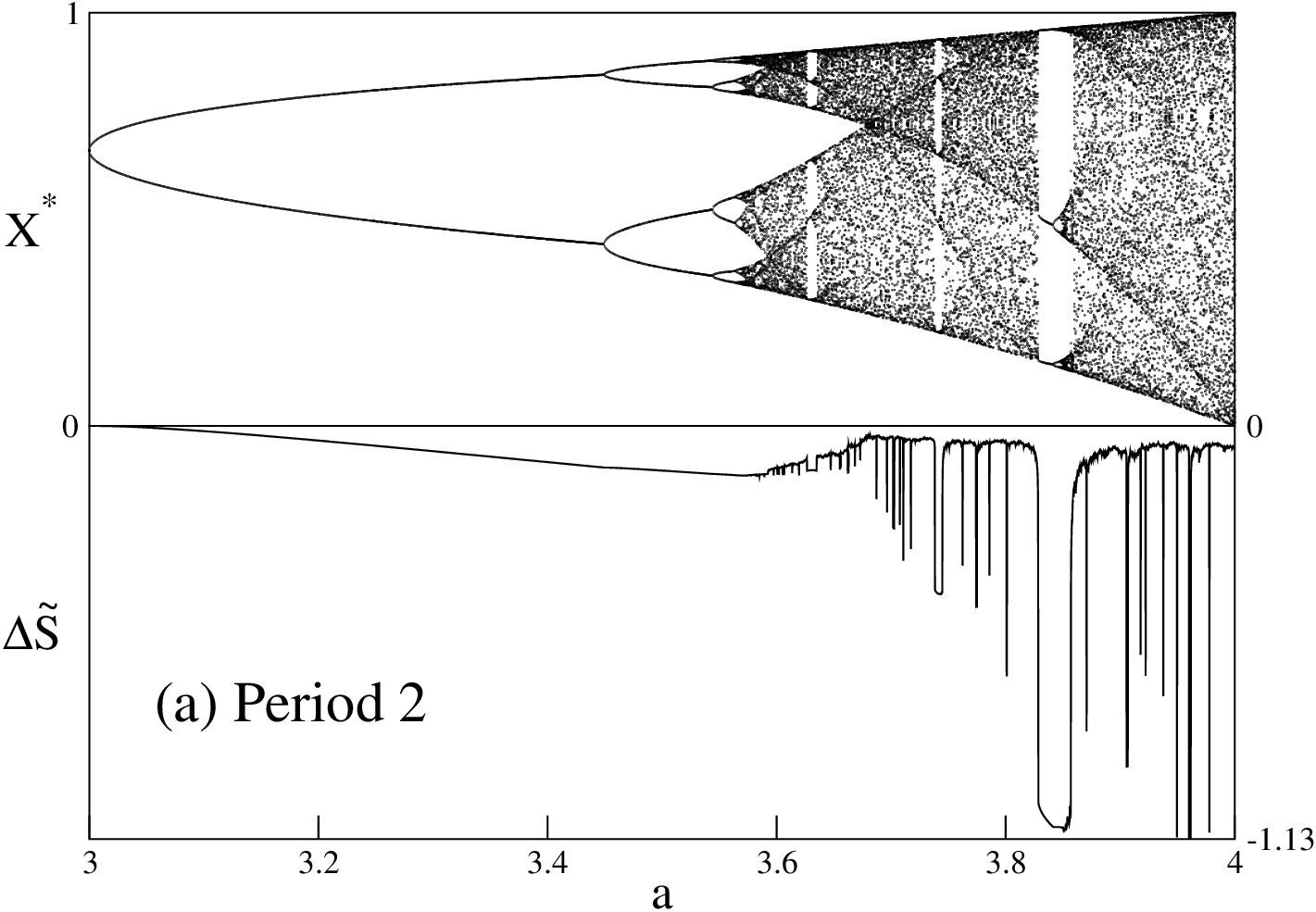}
\includegraphics*[height=6.5cm]{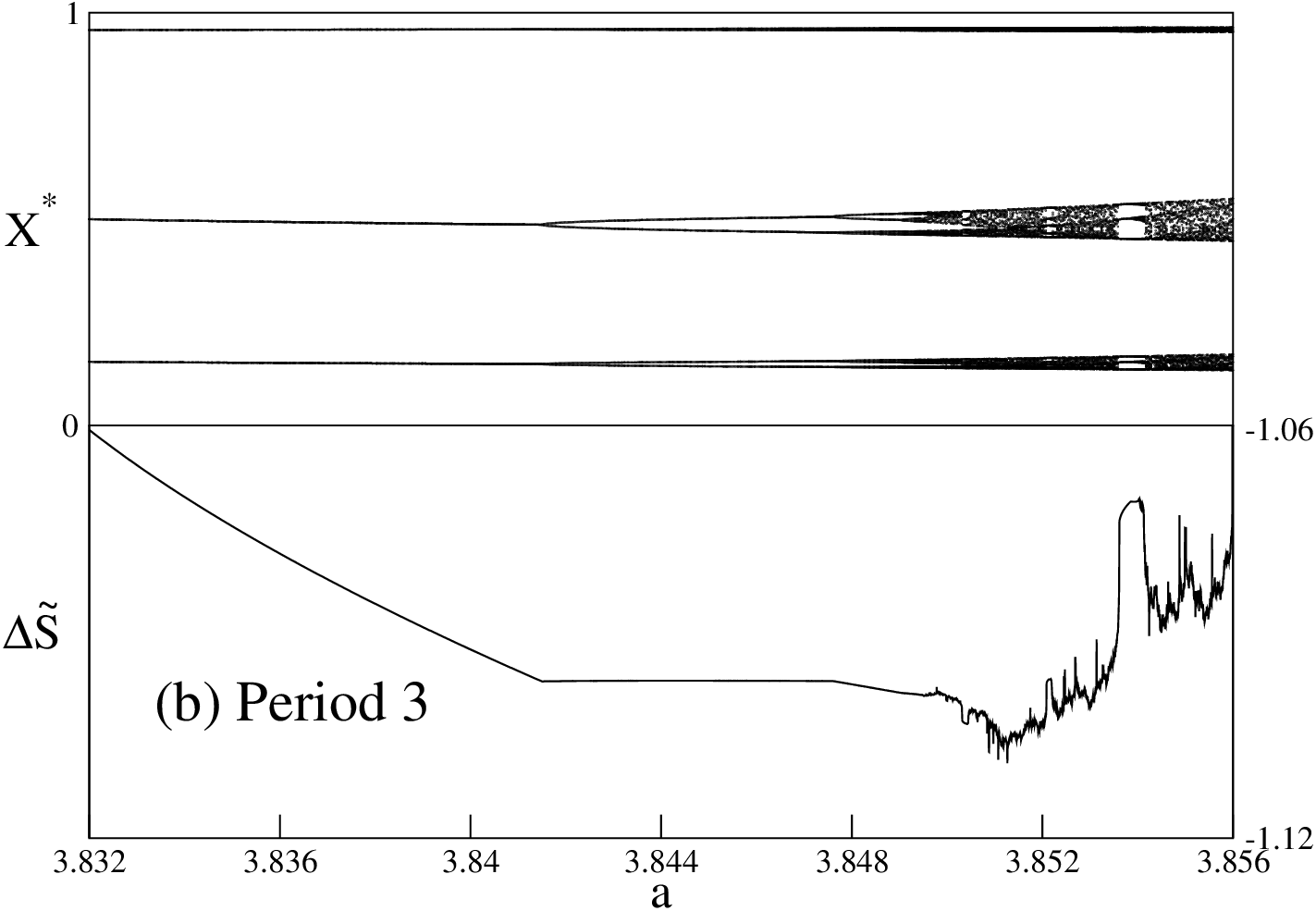}
\includegraphics*[height=6.5cm]{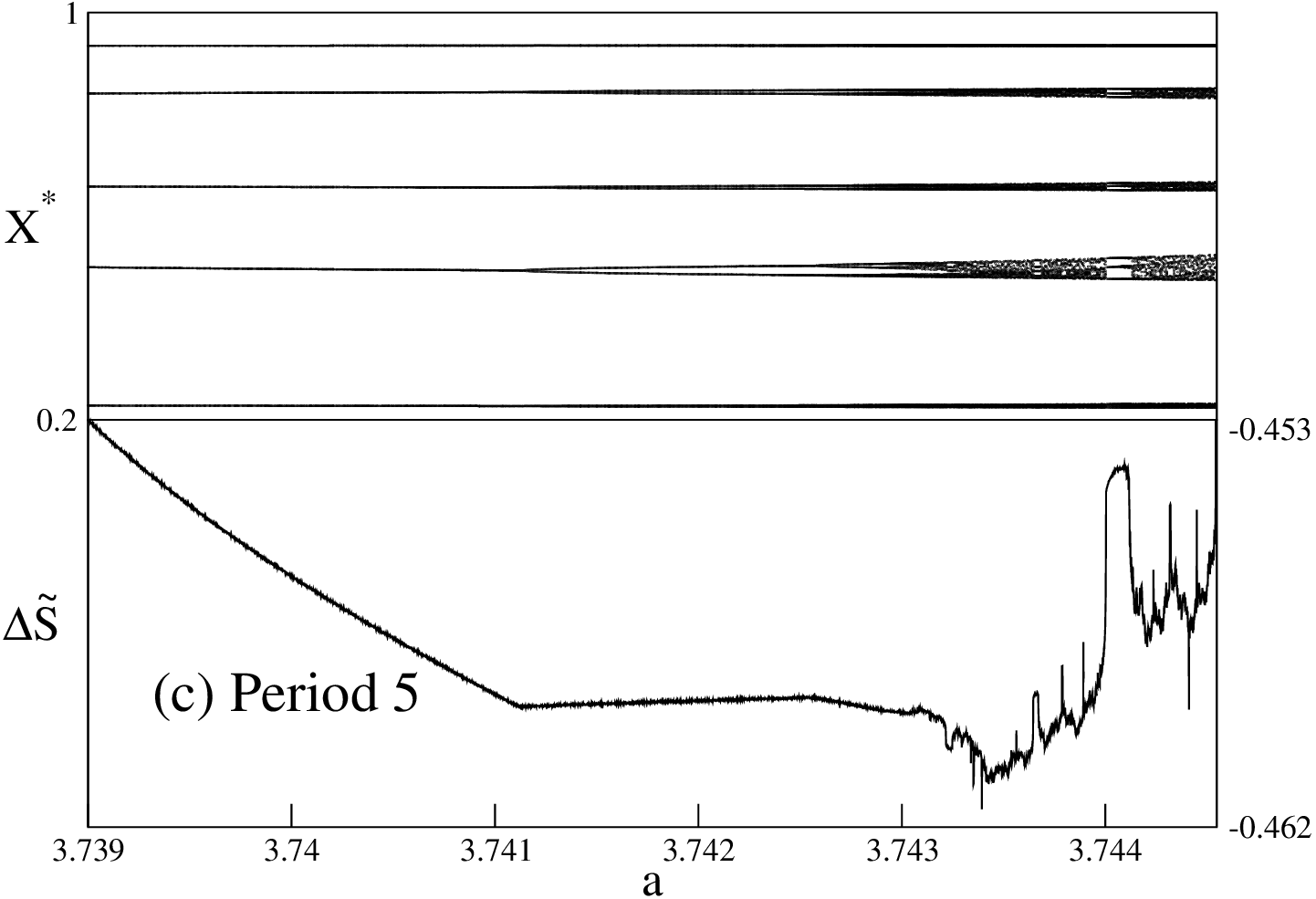}
 \caption{The bifurcation diagram (upper panels) and renormalized entropy
    (lower panels) for period~$2$ (a), period~$3$ (b) and period~$5$ (c)
    of the logistic map.}
\label{fig2c}
\end{center}
\end{figure}

\subsection{Renormalized entropy for the sine-circle map: quasi-periodic route to chaos}

The sine-circle map can exhibit periodic, quasi-periodic or
chaotic behaviors depending on the frequency ratio and the
nonlinearity parameters i.e., $\Omega$ and $K$, respectively. For
$0\leq K\leq 1$, the system dynamics is either periodic
(frequency-locked) or quasi-periodic depending on the value of the
frequency ratio parameter $\Omega$ being rational or irrational.
As the nonlinearity parameter $K$ approaches zero, the
system exhibits quasi-periodic behavior for all values of the
frequency ratio parameter $\Omega$.

As the nonlinearity parameter $K$ approaches one, frequency-locked
steps extend and occupy all $\Omega$ axes where $K$ is equal to one.
In this case, there is a special fraction of $\Omega$ value
called the most irrational $\Omega_{c}$, corresponding to the ``golden
mean'' winding number $W$ if frequency ratio parameter $\Omega$ is
locked to its critical value $\Omega_{c}$. Shortly after this
critical value on $(K,\Omega)$ plane,  $(1,\Omega_{c})$  is the
edge of quasi-periodic route to chaos since chaotic behavior can
occur. All these characteristic shapes on $(K,\Omega)$ plane is
called ``Arnold Tongues'' in the literature. For the $K>1$ region where
the nonlinearity parameter $K$ is dominant on the system dynamics,
there could be periodic regions with different periods, chaotic
regions, and so edges of periodic route to chaos. Also, for this
region, there could be periodic windows possessing same
universality class with the logistic map.

Fig.~2 shows the behavior of the renormalized entropy and the
bifurcations of the sine-circle map for $\Omega_{4}=0.592523777531...$, 
$\Omega_{5}=0.611736528912...$ and $\Omega_{c}=0.606661063469...$
obtained from Eqs.~(\ref{circlemap}-\ref{winding}), 
respectively where the nonlinearity parameter
$K$ lies between zero and $2.5$. The reference state for the
renormalized entropy is chosen to be the one with $K=0$ and
$\Omega=0$ where the system evolves towards a unique stable point.
Note that the degree of irrationality of the system increases as
one moves from Fig.~2a towards Fig.~2c. 

In each of the
aforementioned figures, the oscillatory behavior of the
renormalized entropy is observed when the system is in the
quasi-periodic regime. This oscillatory behavior is exhibited when
$K\in \left[ 0,0.80\right]$, $\left[ 0,0.95\right]$, $\left[
0,1\right]$ for the cases $\Omega$=$\Omega_{4}$, $\Omega_{5}$ and
$\Omega_{c}$, respectively. It is interesting to note that the
interval of the nonlinearity parameter $K$ increases in regard to
the oscillatory behavior of the renormalized entropy as the value
of the frequency ratio parameter reaches the critical value
$\Omega_{c}$, at which the winding ratio $W$ attains the golden
mean. As a result, the renormalized entropy can detect the
quasi-periodic regime as can be seen from Fig.~2. It is worth
noting that the Lyapunov exponent is zero for all quasi-periodic
regions as well as periodic regimes at the bifurcation points
\cite{hilborn}. In this sense, the renormalized entropy is
superior to the Lyapunov exponent, since the renormalized entropy
behaves in a distinct manner in both of the aforementioned
regions. Many chaotic and periodic regions with different periods
are present in Fig.~2 for the nonlinearity parameter $K$ values
$K\in \left[ 0.80,2.50\right]$, $\left[ 0.95,250\right]$, $\left[
1,2.5\right]$ corresponding to $\Omega$=$\Omega_{4}$, $\Omega_{5}$
and $\Omega_{c}$, respectively. The renormalized entropy always
attains values close to zero in these intervals for the chaotic
regions, while it decreases with the increasing number of periods
in the periodic regions until it reaches the edge of chaos. This
can be considered as the signature of the relative degree of order
within the system.

\begin{figure}
\begin{center}
\includegraphics*[height=6.5cm]{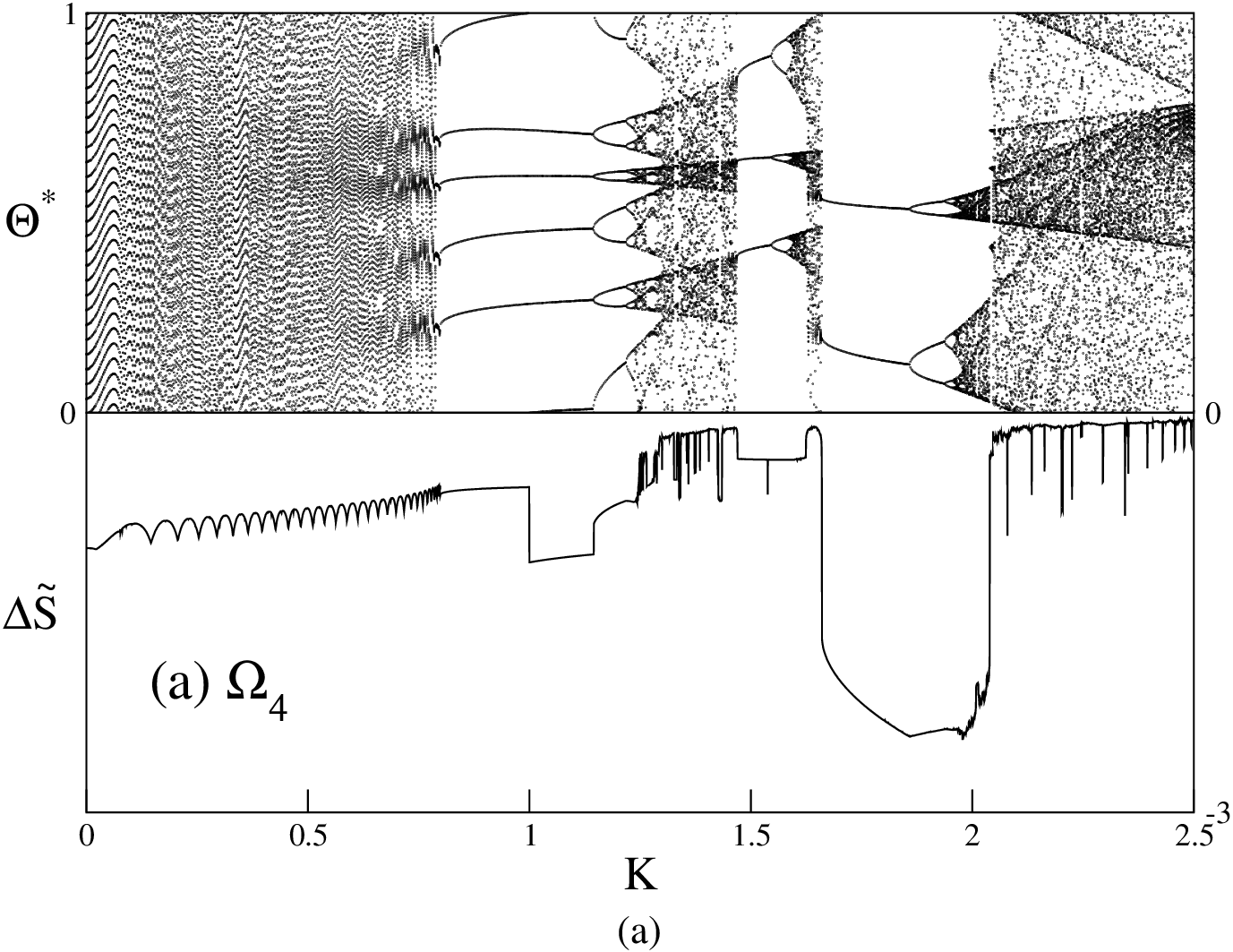}
\includegraphics*[height=6.5cm]{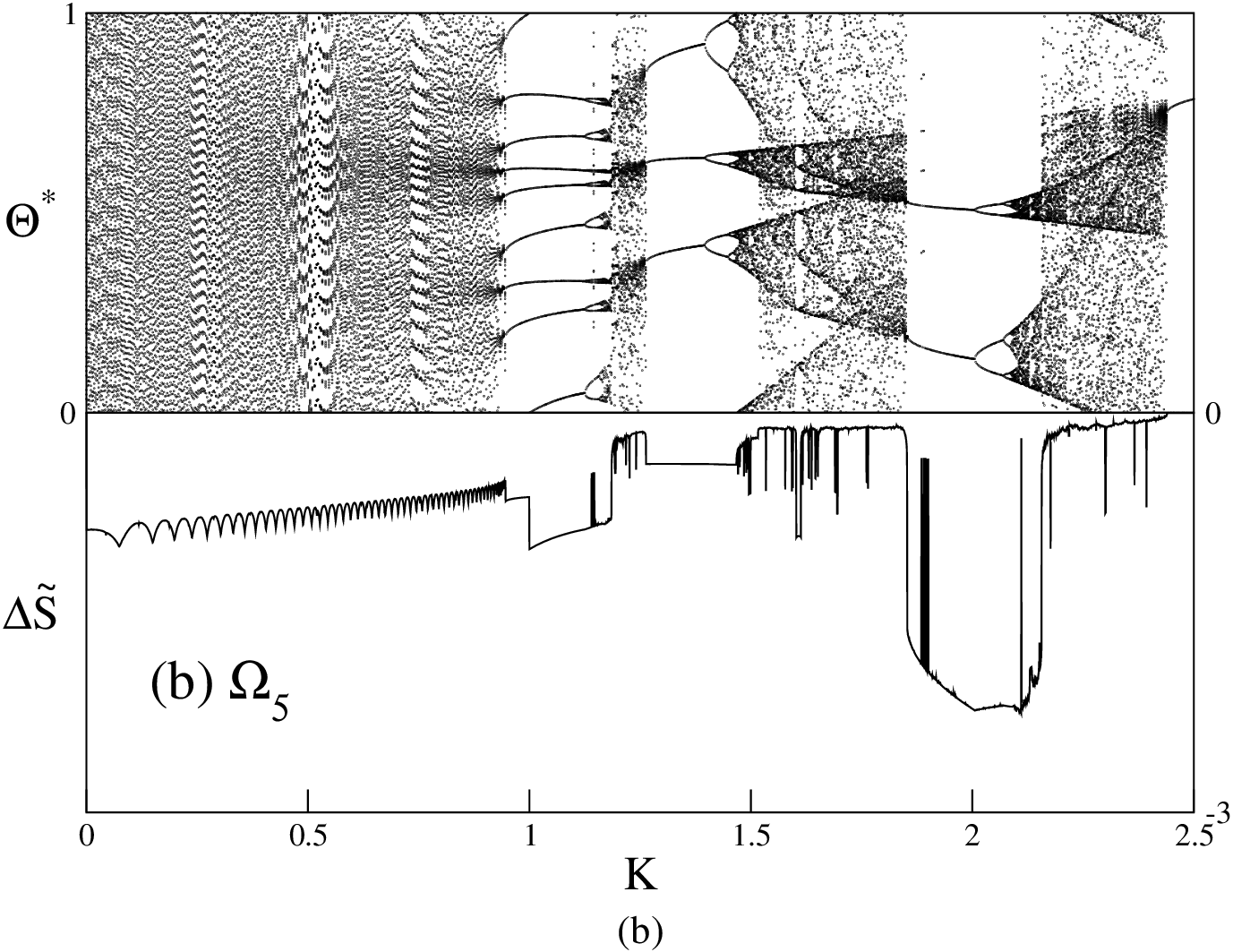}
\includegraphics*[height=6.5cm]{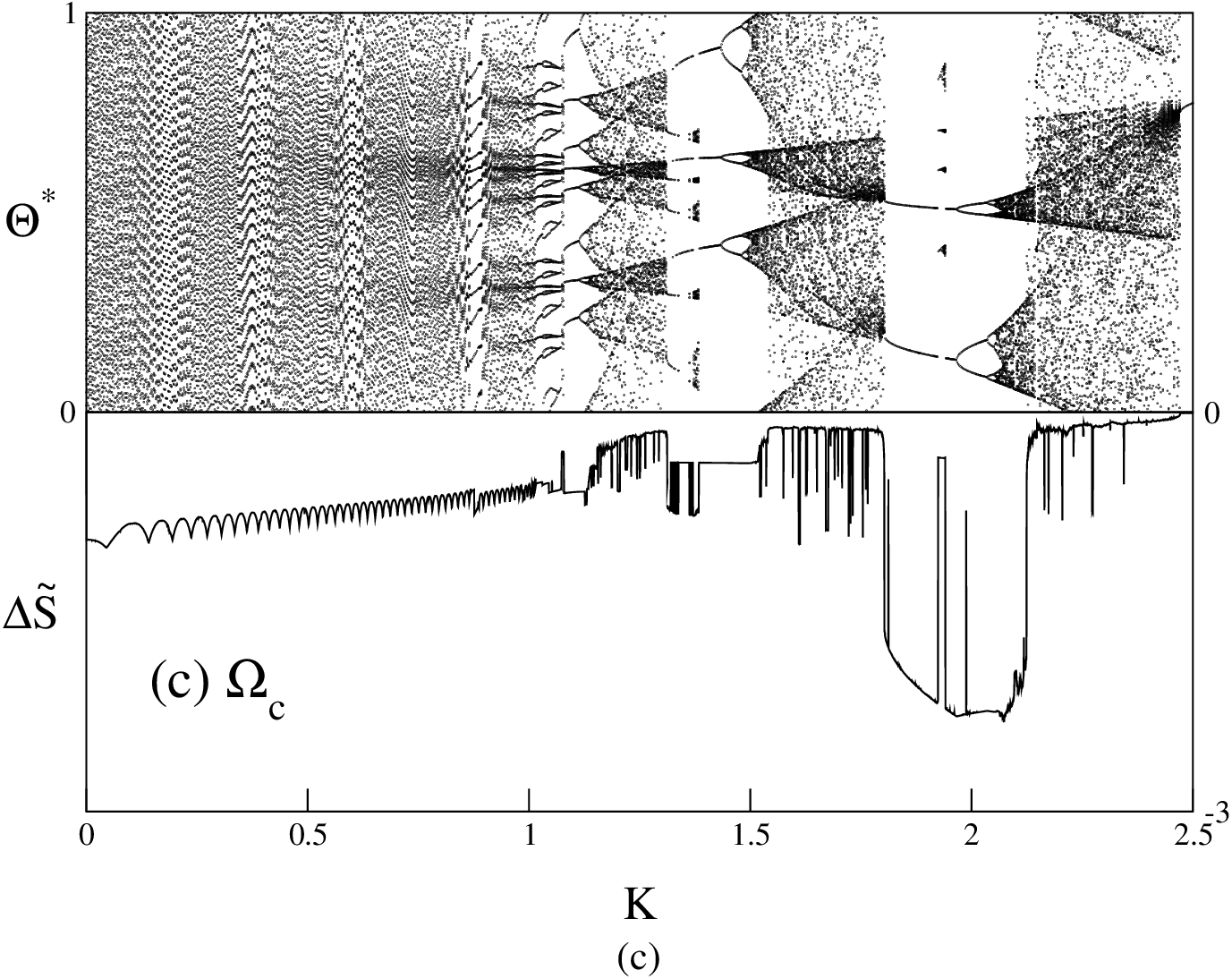}
 \caption{The bifurcation diagram (upper panels) and renormalized entropy
 (lower panels) for $\Omega_4$ (a), $\Omega_5$ (b) and 
 $\Omega_c$ (c) of the sine-circle map.}
\label{fig2c}
\end{center}
\end{figure}

It is well-known that the sine-circle map is in the same
universality class as the logistic map for $K\in \left[
3.19,3.81\right]$ when $\Omega=\Omega_{c}$. Fig.~3 shows the
bifurcation and the renormalized entropy for this particular
window. The renormalized entropy behaves exactly as it does in
Fig.~1a for the logistic map, thereby indicating that different
dynamical maps exhibit same behavior in the regions falling into
the same universality class.

\begin{figure}
\centering
\includegraphics[height=6.5cm,width=10cm]{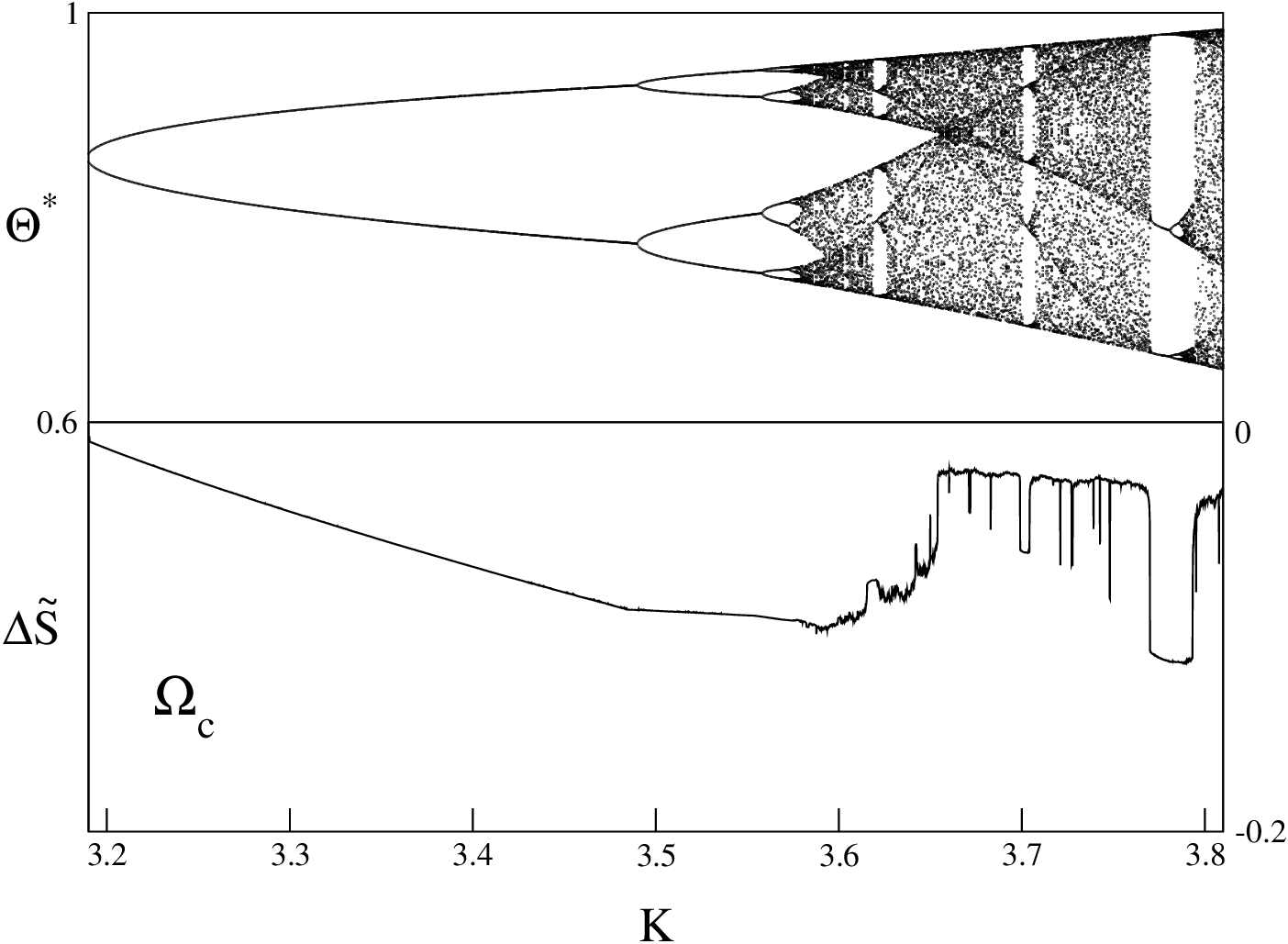}
\caption{The bifurcation diagram (upper panel) and renormalized entropy
(lower panel) of  the region possessing the same universality 
class with the logistic map for $\Omega_c$ of the the sine-circle map.}
\label{fig3}
\end{figure}

\section{Conclusions}

Despite the presence of many different complexity measures, the
ones enabling a local comparison of the distributions are quite
few (for a recent example, see Ref. \cite{romera}). One such
measure of relative nature is the renormalized entropy introduced
by Klimontovich, Kurths and 
coworkers \cite{Kurths1,Kurths2,Kurths3,Kurths4,Klimontovich1,Klimontovich2,Klimontovich3,
Klimontovich4,Klimontovich5,Kopitzki}. 
In this work, the renormalized entropy is used to analyze the
logistic and sine-circle maps. In the former example of the logistic map,
renormalized entropy decreases (increases) up to the accumulation
point (after the accumulation point until the most chaotic state)
as a sign of increasing (decreasing) relative degree of order in all the self-similar 
periodic windows, thereby proving the robustness of this complexity measure. 
By robustness, we emphasize the similarity of the behavior of the renormalized entropy 
in all the self-similar windows, therefore removing the doubt concerning a possible 
accidental feature of the renormalized entropy as a complexity measure. 
On the other hand, The aforementioned observed changes in the renormalized entropy 
are reasonable, since the bifurcations occur before the accumulation point, 
after which the band-merging, in opposition to the bifurcations, is exhibited. 
On top of the precise detection of the accumulation points in all these windows, 
we see that the renormalized entropy can detect the self-similar windows in the
chaotic regime by exhibiting sudden changes in its values.
For the sine-circle map, on the other hand, the renormalized
entropy detects also the quasi-periodic regimes by signaling
oscillatory behavior particularly in these regimes. Moreover, the
oscillatory regime of the renormalized entropy corresponds to a
larger interval of the nonlinearity parameter of the sine-circle
map as the value of the frequency ratio parameter reaches the
critical value, at which the winding ratio attains the golden
mean. Lastly, we remark that the renormalized entropy is
superior to the Lyapunov exponent as a complexity measure, 
since the renormalized entropy can detect the quasi-periodic regimes 
as well as the periodic regimes at the bifurcation points in a distinct 
manner whereas the Lyapunov exponent is zero for both of these regions, 
hence detecting no difference at all \cite{hilborn}.

\section{Acknowlegments}
This work has been supported by TUBITAK (Turkish Agency) under the Research 
Project number 112T083.  U.T. is a member of the Science Academy, Istanbul, Turkey.


\end{document}